\documentclass{article}
\usepackage{graphicx} 
\usepackage[a4paper]{geometry} 
\usepackage{multicol} 
\usepackage{amsmath} 
\usepackage{fancyhdr} 
\usepackage{amsfonts, pdfpages}
\usepackage{amssymb}
\usepackage{makeidx}
\usepackage[ruled,vlined]{algorithm2e}
\usepackage{graphicx}
\usepackage{color}
\usepackage{setspace}
\usepackage{hyperref}
\usepackage{lscape}
\usepackage{float}
\usepackage{comment}

\newtheorem{thm}{Theorem}[section]
\newtheorem{rmk}{Remark}[section]

\newtheorem{lem}{Lemma}[section]

\newtheorem{cor}{Corollary}[section]
\newtheorem{pro}{Proposition}[section]
\newcommand{\e}{\epsilon}
\newcommand{\mbb}{\mathbb}
\newcommand{\mc}{\mathcal}
\newcommand{\mb}{\boldsymbol}
\newcommand{\la}{\lambda}
\newcommand{\mr}{\mathring}
\begin{document}
\title{A vector Allee effect in mosquito dynamics\footnote{The authors were supported by the DSI/NRF SARChI M3B2 grant N 82770\\ GAN acknowledges the Cameroonian Ministry
of Higher Education through the initiative for the modernization of research in Cameroon’s Higher Education }}
\author{J. Banasiak$^{1,2,}$\footnote{Corresponding author jacek.banasiak@up.ac.za},\, Bime M. Ghakanyuy$^{1,4}$\, and Gideon A. Ngwa$^{1,3}$}
\date{\small{1. Department of Mathematics and Applied Mathematics, University of Pretoria, South Africa\\
2. Institute of Mathematics, Łódź University of Technology, Poland\\
3. Department of Mathematics, University of Buea, Cameroon\\
4.  Department of Mathematics and Computer Science, University of Bamenda, Cameroon
}}
\date{\small{The paper is dedicated to Professor Avner Friedman on the occasion of his 90th birthday}}
\maketitle
\begin{abstract}
We consider a recently introduced model of mosquito dynamics that includes mating and progression through breeding, questing and egg-laying stages of mosquitoes using human and other vertebrate sources for blood meals. By exploiting a multiscale character of the model and recent results on their uniform-in-time asymptotics, we derive a simplified monotone model with the same long-term dynamics.  Using the theory of monotone dynamical systems, we show that for a range of parameters, the latter displays Allee-type dynamics; that is, it has one extinction and two positive equilibria ordered with respect to the positive cone $\mbb R_+^7$, with the extinction and the larger equilibrium being attractive and the middle one unstable. Using asymptotic analysis, we show that the original system also displays this pattern.
\end{abstract}

\section{Introduction}
Understanding the life cycle of mosquitoes is of utmost importance in any attempt to control diseases caused by them, such as malaria, dengue, zika or chikungunya, see, e.g., \cite{NMBB,ghakanyuy2022investigating,dumont2016human}. The problem is that it is pretty complex. Broadly, the cycle can be divided into the aquatic and terrestrial stages. In the aquatic stage, we have eggs, several larval stages and pupae, which hatch into terrestrial juvenile male and female mosquitoes who mate, and the female fertilized (breeding) mosquitoes quest for blood necessary for the development of eggs. The fed female mosquitoes lay eggs and return to the breeding stage. There are various ways of modelling the cycle at different levels of resolution, focusing on its particular parts. Many models have provided a simplified picture of the cycle. For instance, in \cite{anguelov2012mathematical,li}, the authors consider male and female mosquitoes but disregard the mating and questing processes. On the other hand, in modelling mosquito-borne diseases, it is essential to recognize that a female mosquito must bite two times, an infected and then a susceptible human, to transmit the disease. Thus, paying closer attention to the gonotrophic cycle between the breeding and ovipositing stages becomes important; see \cite{ngwa2019three}.
This paper will consider the model introduced in \cite{NMBB}, whose flowchart is presented in Fig. \ref{fmod1}. The complexity of the model and lack of a clear structure makes its comprehensive analysis difficult; in fact, \cite{NMBB} contains only a preliminary description of its dynamics. On the other hand, the model is driven by several processes with widely different rates. For instance, mating occurs quickly after the emergence of female mosquitoes to ensure the passage of their genes to the next generation, \cite{VDCI}. Consequently, the dynamics of juvenile mosquitoes can be considered to be much faster than that of aquatic and adult forms. Also, there is evidence, \cite{clements1992biology,goindin2015parity,VDCI}, that the life span of newly hatched male mosquitoes is much shorter than that of the females. These factors motivate the possibility of considering the system as a multiscale one and exploiting this property to simplify it using the methods of the singular perturbation theory, \cite{BL,Kue,Hek}, as done in \cite{BanTch,CapMBS,RashKooi,RashVent,RASS}.   Moreover, by more refined techniques of \cite{Hopp, BanViet}, it is possible to show that the simplified models have the same long-term dynamics as the original one.

The paper has the following structure. In Section \ref{sect1}, we introduce the model and discuss its multiscale versions and resulting simplified systems, which turn out to be cooperative, allowing for a reasonably complete description of their long-term dynamics. We focus on the case with a large death rate of the juvenile males and provide a detailed description of its dynamics in Section \ref{sec3}. In particular, we show the existence of a saddle-node bifurcation of equilibria. An interesting feature here is that the bifurcation leads to the creation of three equilibria, the trivial and two positive ones, which, moreover, are ordered (with respect to the positive orthant of the state space). Furthermore, the trivial and the largest equilibria are globally asymptotically stable in adjoining order intervals. In contrast, the middle equilibrium is unstable, which shows that the model exhibits a vector Allee effect. In Section \ref{sec4}, we apply the results of \cite{BanViet,BanM2AS} to discuss the relevance of the analysis of the simplified model to the long-term dynamics of the original one and in Section \ref{sec5}, we present simulations showing, in particular, that the former provides a good approximation to the latter even for relatively small differences between the juvenile male and female mosquitoes death rates. Finally, Section \ref{sec6} contains conclusions.

\begin{table}
\begin{tabular}{|p{2.8cm}|p{10cm}|}
\hline \textbf{State Variables} &  \textbf{Description of Variables}  \\
\hline $A$ &   Density of  aquatic life forms \\
\hline $F_B$ & Density of newly emerged (unfertilized) female  mosquitoes  \\
\hline $M_B$ & Density of newly emerged  male mosquitoes \\
\hline $B$ & Density of fertilized  female mosquitoes \\
\hline $Q_{H}$ & Density of fertilized  female mosquitoes questing human blood \\
\hline $Q_{V}$ & Density of fertilized female mosquitoes seeking non human blood  \\
\hline $R_{V}$ & Density of  mosquitoes that successfully fed from  nonhuman
source  \\
\hline $R_{H}$ &  Density of  mosquitoes that successfully fed from humans
 \\
\hline $H$ & Constant human population density\\
\hline $V$ & Constant vertebrate population density \\
\hline
\end{tabular}
\caption{Description of state variables}
  \label{table1}
\end{table}

\section{The model and its asymptotic reduction}\label{sect1}

\begin{figure}[H]
\centering
\includegraphics[scale=0.8]{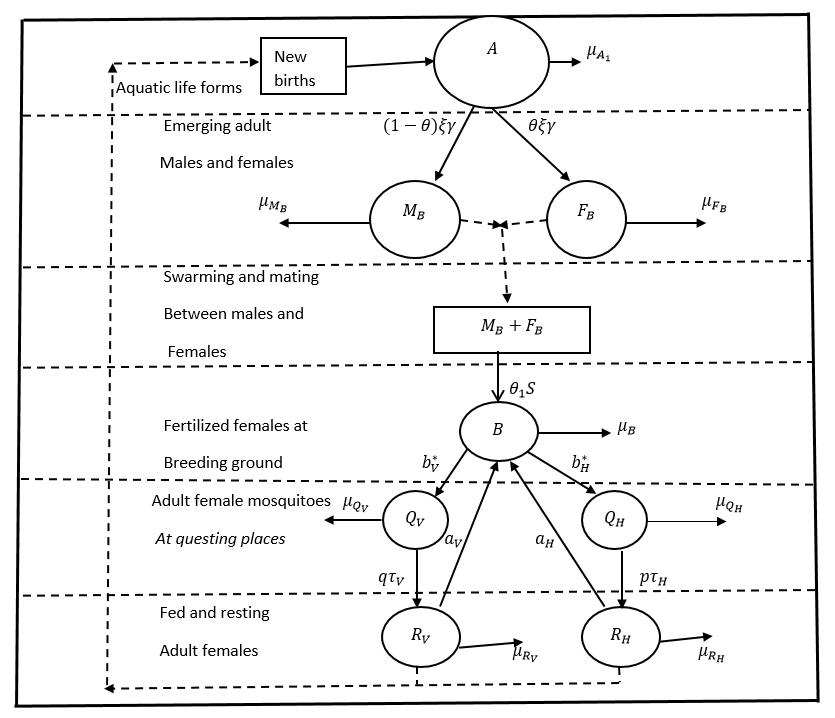}
\caption{Compartments of the model.} \label{fmod1}
\end{figure}

To describe the model, we introduce the state variables in Table \ref{table1} and the interactions between them in the flow chart in Figure \ref{fmod1}.
Thus, the original model from \cite{NMBB} can be written as
\begin{equation}\label{eq:fullsystem}
\begin{split}
A' & = (a_{H}R_{H}+a_{V}R_{V}) \lambda(R) \left(1-\frac{A}{L_P}\right) - \left(\gamma+\mu_{A1}+\mu_{A2}A \right)A,\\ 
M_B'& =  (1-\hat\theta) \hat\xi \gamma A-\hat\mu_{M_B}M_B, \\ 
F_B' & =  \hat\theta  \hat\xi \gamma A - SM_BF_B-\hat\mu_{F_B} F_B,\\ 
B' & =  SM_BF_B+a_{H}R_{H}+a_{V}R_{V}-bB-\mu_{B}B, \\ 
Q_{H}' & =  b\omega B -\tau_{H}HQ_{H}-\mu_{Q_{H}}Q_{H},\\ 
Q_{V}' & =  b(1-\omega)B -\tau_{V} V Q_{V}-\mu_{Q_{V}}Q_{V},\\ 
R_{H}' & =  p\tau_{H}HQ_{H}-a_{H}R_{H}-\mu_{R_{H}}R_{H},\\ 
R_{V}' & = 
q\tau_{V}V Q_{V}-a_{V}R_{V}-\mu_{R_{V}}R_{V},
\end{split}
\end{equation}
where $R=R_{V}+R_{H}$. Most terms are self-explanatory with the rates described in Table \ref{table2}.
 Transfers between stages are linear except for the oviposition rate, the density-dependent death rate of the aquatic stage, see \cite{anguelov2012mathematical}, and the mating rate, modelled using the mass action law. In the oviposition rate, given by \begin{equation}
\Lambda(R_H,R_V)\left(1-\frac{A}{L_P}\right)=(a_{H}R_{H}+a_{V}R_{V}) \lambda(R) \left(1-\frac{A}{L_P}\right),\label{La}
\end{equation}
$\la(R)$ is the change in the oviposition per capita due to overcrowding. A comprehensive description of the function $\la$ can be found in \cite{NMBB}. In general, if it is defined for all $R$, it should be a non-increasing function for large $R$ and such that $\Lambda(R_H,R_V)$ has a finite limit as $R\to \infty$. It is, however, possible that $\la$ is only defined for some interval $[0,L]$ if a certain level of overcrowding totally prevents the female mosquitoes from laying eggs, \cite{NMBB}. Similarly, the factor  $\left(1-\frac{A}{L_P}\right)$  is  specific to some mosquito species. E.g.,  \textit{Aedes albopictus} (and even \textit{Aedes aegypti}) are capable of selecting their breeding sites, seeking those with high food content and low intraspecific competition pressure. Thus, if breeding sites in a given area already contain too many larvae, the females will not deposit eggs there, see \cite{dumont2016human}. The only parameters that require explanation are $\gamma$ and $\xi$ (and $\hat\xi)$. The need to introduce both follows from the fact that the aquatic stage is described by its biomass, while $M_B$ and $F_B$ are the number of individuals (or densities). Thus, $\gamma$ is the rate of depletion of the aquatic biomass by the hatching of juvenile mosquitoes, and $\xi$ is the conversion factor of the unit of aquatic biomass into individual mosquitoes.     We note that in \eqref{eq:fullsystem}, we use "hatted" notation $\hat\xi,\hat\mu_{M_B}, \hat\mu_{F_B}$ and $\hat\theta$ to indicate the parameters which can be considered large (or small) in asymptotic analysis.

To complete the formulation of the above system, we complement \eqref{eq:fullsystem}  with initial conditions
\begin{equation}\label{eq:InitCond}
\begin{split}
A(0) & = \mr A,~M_B(0)=\mr M_{B},~~F_B(0)=\mr F_{B},~~R_{V}(0) = \mr R_{V},~~R_{H}(0)=\mr R_{H},\\
B(0) & =\mr  B,~~Q_{V}(0)=\mr Q_{V},~~Q_{H}(0)=\mr Q_{H},
\end{split}
\end{equation}
where $\mr A,\ldots, \mr Q_{H} $ are all non-negative. 
\begin{table}
\begin{tabular}{|p{2cm}|p{11.5cm}|}
\hline
\textbf{Parameter} & \textbf{Description of Parameters}  \\
\hline $\theta, \hat\theta$ & Proportion of aquatic forms that develop
into female mosquitoes.  \\
\hline $\gamma$ & Rate of the aquatic forms transition into terrestrial forms.  \\
\hline $\xi, \hat{\xi}$ & Biomass transition factor from aquatic into  terrestrial forms.  \\
\hline  $\mu_{y},\hat{\mu}_y$ & Natural death rate of the mosquitoes of type $y$, where $y$ denotes one
of the adult mosquito-types.   \\
\hline $a_{V},\: a_{H}$ & Return rates of reproductive females to the breeding site from to the
 from, respectively,  vertebrates ($a_{V}$) and human ($a_{H}$) habitats.\\
\hline
$b$ & Rate at which mosquitoes leave the breeding site for questing.\\
\hline $\omega$ & The human blood preference factor.  \\
\hline $\tau_{V}$ & Effective mass action contact parameter between
zoophilic mosquitoes and vertebrates. \\
\hline $\tau_{H}$ & Effective mass action contact parameter between
anthropophilic mosquitoes and humans.  \\
\hline $p,q$ & Probabilities that a mosquito successfully harvests a blood meal from a
human ($p$) or vertebrate ($q$) population.  $0\leq p,q \leq 1.$\\
\hline $\mu_{A1}$ & Natural death rate per aquatic life form. \\
\hline $\mu_{A2}$ & Additional death due to overcrowding of aquatic forms\\
\hline $S$ & Constant mass action contact parameters between female and male mosquitoes. \\
\hline $L_P$ & Carrying capacity of the pond.  \\
\hline
\end{tabular}
\caption{Table of parameters and their descriptions. The "hatted" parameters are rescaled in the process of multiscale analysis, and the "unhatted" versions are the reference values.  } \label{table2}
\end{table}

\subsection{Asymptotic reduction}
In this section, we formally simplify system \eqref{eq:fullsystem} using an asymptotic approach based on the existence of multiple time scales in it. Then, the reduced systems allow for a detailed mathematical analysis, which can be translated back to the level of  \eqref{eq:fullsystem}. To identify the different time scales, we refer to biological considerations. 

\noindent
1. There is evidence, \cite{clements1992biology,goindin2015parity,VDCI}, that the life span of newly hatched male mosquitoes $M_B$ is much shorter than that of the females $F_B,$ that is, $\hat\mu_{M_B}$ is much larger than $\hat\mu_{F_B}.$ To ensure that despite the high mortality of the males, their population does not vanish, the recruitment rate of the males must be of a comparable magnitude. Since $\hat\theta \in (0,1)$ and making $\gamma$ large would drive $A$ to extinction, the only possibility is that $\hat\xi$ is also large. However, if we want to keep $\mu_{F_B}$ and $S$ at the slow-time scale, the recruitment of $F_B$ must also be slow, which yields that $\hat\theta$ must be small. Thus, we re-label the parameters as $\e\hat\mu_{M_B}=\mu_{M_B}$, $\e\hat\xi=\xi$ and $\hat\theta=\e\theta,$ where $\e$ is a small positive parameter and $\mu_{M_B},\xi, \theta$ are $O(1)$ quantities.

This argument results in the singularly perturbed system
\begin{equation}\label{eq:fullsystem1}
\begin{split}
A' & = \Lambda(R_H,R_V) \left(1-\frac{A}{L_P}\right) - \left(\gamma+\mu_{A1}+\mu_{A2}A \right)A,\\ 
\e M_B' & =  (1-\e\theta) \xi \gamma A-\mu_{M_B}M_B, \\ 
F_B' & = \theta \xi \gamma A - SM_BF_B-\mu_{F_B} F_B,\\ 
B' & = SM_BF_B+a_{H}R_{H}+a_{V}R_{V}-bB-\mu_{B}B, \\ 
Q_{H}' & =  b\omega B -\tau_{H}HQ_{H}-\mu_{Q_{H}}Q_{H},\\ 
Q_{V}' & =  b(1-\omega)B -\tau_{V} V Q_{V}-\mu_{Q_{V}}Q_{V},\\
R_{H}' & =  p\tau_{H}HQ_{H}-a_{H}R_{H}-\mu_{R_{H}}R_{H},\\ 
R_{V}' & =  q\tau_{V}V Q_{V}-a_{V}R_{V}-\mu_{R_{V}}R_{V}.
\end{split}
\end{equation}
Note that \eqref{eq:fullsystem1} is the same as \eqref{eq:fullsystem} if $\e=\frac{\mu_{M_B}}{\hat\mu_{M_B}}=\frac{\xi}{\hat\xi}=\frac{\hat\theta}{\theta},$ so $\e$ is small if $\hat\mu_{M_B}$ and $\hat\xi$ are large and $\hat\theta$ is small. We emphasize that this does not place any restriction on $\e$ as we have the freedom of choosing $\mu_B,\xi$ and $\theta$ as long as they are of the same magnitude as the other parameters. 

Letting $\e=0$, we arrive at the equation of the slow manifold (quasi-steady state), \cite{BL,Kue,Hek},
\begin{equation}
 M_B = \frac{\xi \gamma A}{\mu_{M_B}},
 \label{qss1}
 \end{equation}
and the  reduced model
\begin{equation}\label{eq:fullsystemmodel1}
\begin{split}
A' & = \Lambda(R_H,R_V) \left(1-\frac{A}{L_P}\right) - \left(\gamma+\mu_{A1}+\mu_{A2}A \right)A,\\  
F_B' & =  \theta \xi \gamma A - \frac{S\xi \gamma}{\mu_{M_B}}AF_B-\mu_{F_B} F_B,\\ 
B' & =  \frac{S\xi \gamma}{\mu_{M_B}}AF_B+a_{H}R_{H}+a_{V}R_{V}-bB-\mu_{B}B, \\ 
Q_{H}' & =  b\omega B -\tau_{H}HQ_{H}-\mu_{Q_{H}}Q_{H},\\ 
Q_{V}' & =  b(1-\omega) B -\tau_{V} V Q_{V}-\mu_{Q_{V}}Q_{V},\\ 
R_{H}' & =  p\tau_{H}HQ_{H}-a_{H}R_{H}-\mu_{R_{H}}R_{H},\\
R'_{V} & =  q\tau_{V}V Q_{V}-a_{V}R_{V}-\mu_{R_{V}}R_{V}.
\end{split}
\end{equation}
2. The other natural assumption is that male and young female dynamics are fast. We argue, as in the previous case, that to prevent both populations from becoming extinct, we need to have also fast recruitment rates, so we assume that $\xi$ is large, but here, there is no need to change the sex differentiation factor $\theta$. Thus, we re-label the parameters as  $\e\hat\mu_{M_B}=\mu_{M_B}$, $\e\hat\mu_{F_B}=\mu_{F_B}$ and  $\e\hat\xi=\xi,$  where $\e$ is a small positive parameter, $\hat\theta =\theta$ and $\theta, \mu_{M_B}, \mu_{F_B}$ and $\xi$ are $O(1).$
Hence, the slow manifold is given by
\begin{equation}
 M_B = \frac{(1-\theta)\xi \gamma A}{\mu_{M_B}},\quad  F_B = \frac{\theta\xi \gamma A}{\mu_{F_B}}
 \label{qss2}
 \end{equation}
 and the reduced system is 
\begin{equation}\label{eq:simplifysystemmodel2}
\begin{split}
A' & = \Lambda(R_H,R_V)\left(1-\frac{A}{L_P}\right)- \left(\gamma+\mu_{A1}+\mu_{A2}A \right)A,\\ 
B' & =  \frac{S\theta \left( 1-\theta\right) \xi^2\gamma^2 }{\mu_{M_B}\mu_{F_B}} A^2+a_{H}R_{H}+a_{V}R_{V}-bB-\mu_{B}B, \\ 
Q_{H}' & =  b\omega B -\tau_{H}HQ_{H}-\mu_{Q_{H}}Q_{H},\\ 
Q_{V}'& =  b(1-\omega)B -\tau_{V} V Q_{V}-\mu_{Q_{V}}Q_{V},\\ 
R_{H}'& =  p\tau_{H}HQ_{H}-a_{H}R_{H}-\mu_{R_{H}}R_{H},\\ 
R_{V}' & =  q\tau_{V}V Q_{V}-a_{V}R_{V}-\mu_{R_{V}}R_{V}.
\end{split}
\end{equation}
We observe that \eqref{qss1} amounts to a biologically reasonable assumption that the newly emerging males are proportional to the existing biomass at a given time. Similarly, \eqref{qss2} translates into the assumption that females and males hatch in the numbers proportional to the current aquatic biomass. The advantage of the multiscale approach is that it explicitly gives us the proportionality constants.  We also note that we can exploit other possible fast processes, e.g., assuming the additional high female death rate due to the dangers of mating, which leads to even smaller systems with similar dynamics. Such models, which can also be derived from first principles, are discussed in \cite{BBN}. Here, we shall focus on   \eqref{qss1} and \eqref{eq:fullsystemmodel1} as the analysis of \eqref{qss2}, \eqref{eq:simplifysystemmodel2} can be carried out in the same way.

\section{Analysis of \eqref{eq:fullsystemmodel1}}\label{sec3}

In this section, we shall analyse the dynamics of \eqref{eq:fullsystemmodel1}. As our approach uses extensively the theory of monotone systems, we begin with the relevant notation. 
We denote open and closed positive cones in $\mbb R^7$ by 
    $$
    \mbb R^n_+ = (0,\infty)^n, \quad \overline{\mbb R}^n_+ = [0,\infty)^n.
    $$
    The latter introduces a partial order in $\mbb R^n$. Accordingly, for any $\boldsymbol{a},\boldsymbol{b}\in \mathbb{R}^n$, we write (i) $\boldsymbol{a}\le \boldsymbol{b}$ when $a_i \le b_i$, (ii) $\boldsymbol{a}< \boldsymbol{b}$ when $a_i \le b_i$ and $\boldsymbol{a}\ne \boldsymbol{b}$, (iii)
	$\boldsymbol{a}\ll \boldsymbol{b}$ when $a_i<b_i$, where $i=1,\dots, n$. For  $\boldsymbol{a}\le \boldsymbol{b}$, we introduce order intervals 
    $$
     [\boldsymbol{a},\boldsymbol{b}]= \{\boldsymbol{x}\in \mathbb{R}^n: \boldsymbol{a}\le \boldsymbol{x}\le \boldsymbol{b}  \},
    $$	
    and for
    $\boldsymbol{a}\ll \boldsymbol{b}$,
    $$
     [\boldsymbol{a},\boldsymbol{b})= \{\boldsymbol{x}\in \mathbb{R}^n: \boldsymbol{a}\le \boldsymbol{x}\ll \boldsymbol{b}  \}.
    $$	
   For future use, denote 
\begin{equation}\label{notation}
\begin{split}
\boldsymbol{x}&=\left (A,F_B,B,Q_H,Q_V,R_H,R_V \right ),\quad \mr{\boldsymbol{x}}=\left (\mr A,\mr F_{B},\mr B,\mr Q_{H},\mr Q_{V},\mr R_{H},\mr R_{V} \right )\\ 
\alpha &= \frac{S\xi \gamma}{\mu_{M_B}},\quad l= L_P^{-1},\quad \sigma_H = \tau_H H, \quad\sigma_V = \tau_V V,\quad T_B= b+\mu _B,\\
T_{Q_H}&=\mu _{Q_H}+\sigma_H,\quad  T_{Q_V}=\mu _{Q_V}+\sigma_V,\quad T_{R_H}=\mu _{R_H}+a_H,\quad T_{R_V}=\mu _{R_V}+a_V.
\end{split}
\end{equation}

\begin{lem}
Consider the system represented by \eqref{eq:fullsystemmodel1} and let $\lambda:[0,\infty)\to [0,\infty)$ be any oviposition rate. Then the flow of \eqref{eq:fullsystemmodel1} is positively invariant on the set
\begin{equation}\label{set:omega}
    \Omega= \left \{ \boldsymbol{x} \in \overline{\mathbb{R}}^7_+: 0\le A \le L_P,~~0\le F_B\le \frac{\theta \mu_{M_B}}{S} \right \}.
\end{equation}
\end{lem}
\noindent \textbf{Proof}. The positivity follows as the off-diagonal terms in \eqref{eq:fullsystemmodel1} are nonnegative for nonnegative $\mb x$, \cite[Theorem B.7]{smith1995theory}. The other bounds follow from the fact that $A'\left(L_P \right)<0$ and $F_B'\left(\frac{\theta  \mu_B }{S} \right)<0$. \hfill $\square$

\begin{thm}\label{thm:cooperativemodel1}
Assume that the partial derivatives $\Lambda_{R_H}(R_H,R_Q)$ and $\Lambda_{R_Q}(R_H,R_Q)$ are nonnegative on $\overline{\mbb R}_+^2$. Then \eqref{eq:fullsystemmodel1} is a cooperative system on $\Omega$.
\end{thm}
\noindent \textbf{Proof}. 
Let 
\begin{equation}\label{eq:jacobianmodel1}
 \mc J (\mb x)= \left(
\begin{array}{ccccccc}
 J_{11} & 0 & 0 & 0 & 0 & \left(1-lA\right)\Lambda_{R_H} & \left(1-lA\right)\Lambda_{R_V}\\
 J_{21} & J_{22} & 0 & 0 & 0 & 0 & 0 \\
 \alpha  F_B & \alpha A  & -T_B & 0 & 0 & a_H & a_V \\
 0 & 0 & b\omega & -T_{Q_H} & 0 & 0 & 0 \\
 0 & 0 & b(1-\omega) & 0 & -T_{Q_V} & 0 & 0 \\
 0 & 0 & 0 & p \sigma_H  & 0 & -T_{R_H} & 0 \\
 0 & 0 & 0 & 0 & q \sigma_V & 0 & -T_{R_V} \\
\end{array}
\right),
\end{equation}
where
\begin{equation} 
    J_{11} =-\mu _{\text{A1}}-\gamma-2 A \mu _{\text{A2}}-l \Lambda(R_H,R_V),\quad J_{21}=\gamma  \xi \theta -\alpha F_B,\quad J_{22}= -\alpha A-\mu _{F_B},\label{eq:bh}
\end{equation}
be  the Jacobian matrix of \eqref{eq:fullsystemmodel1}. By \eqref{set:omega} and the assumption, $J_{21}\geq 0,$ $\left(1-lA\right)\Lambda_{R_H}\geq 0$ and $\left(1-lA\right)\Lambda_{R_V}\geq 0$, so that $\mc J(\mb x)$
is a Metzler matrix on $\Omega$. \hfil $\square$
\begin{rmk}
The typical forms of $\la$ satisfying the assumption of Theorem \ref{thm:cooperativemodel1} are $\la(R) = \la_0$ for some constant $\la_0$, see \cite{dumont2016human, anguelov2012mathematical, ngwa2019three} or a Holing type function with $a_H=a_V=a$
$$
\Lambda(R) = \la_0a\frac{R}{c+R}.
$$
Indeed, for $i=H,V$,
$$
\Lambda_{R_i}(R_H,R_V) = \la_0\frac{ac}{(c+R)^2}\geq 0.
$$
In the first case, the per capita oviposition rate is not affected by the overcrowding, the growth of the resulting aquatic forms is constrained by the pond's carrying capacity so the aquatic population cannot exceed $L_P$ irrespective of the number of ovipositing females. On the other hand, the Holling type rate models the situation when the average per capita oviposition rate decreases with overcrowding, but the total number of laid eggs does not decrease to zero. 
\end{rmk}
For the rest of this work, we shall assume
that the mosquito population belongs to $\Omega$. To proceed with the analysis, we assume that 
\begin{equation}
\Lambda (R_H,R_A)=\lambda_0(a_HR_H+a_VR_V).
\label{Laass}
\end{equation}
Next, we consider the existence of equilibria. The equation for equilibria can be written as
\begin{equation}
\begin{split}
0& = \Lambda(R_H,R_V) \left(1-l{A}\right) - \left(\gamma+\mu_{A1}+\mu_{A2}A \right)A, \\
	0& =  \theta \xi \gamma A - \alpha AF_B-\mu_{F_B} F_B,\\
0& = \mathcal{L}\boldsymbol{y}+\boldsymbol{G},
\label{Yeq}
\end{split}
\end{equation}
where
$$
 \mathcal{L}\! =\! \left(\!\!\begin{array}{ccccc}
-T_{B}&0&0&a_{H}&a_{V}\\
 b\omega& -T_{Q_H}&0&0&0\\
 b(1-\omega)&0&-T_{Q_V} &0&0\\
0& p\sigma_H&0&-T_{R_{H}}&0\\
0&0& q\sigma_{V}&0& -T_{R_{V}}\end{array}\!\!\right),\quad 
\boldsymbol{y}=\left(\!\!\begin{array}{c}B\\Q_H\\Q_V\\R_H\\R_H\end{array}\!\!\right),\quad  \boldsymbol{G}=\left(\!\!\begin{array}{c}\alpha AF_B\\0\\0\\0\\0\end{array}\!\!\right).$$
We have
$$F_B^*(A) = \frac{\theta \xi \gamma A}{\mu_{F_B} +  \alpha A}
$$
and, using the fact that $\mc L$ is a nonsigular Metzler matrix,
$$
\mb y^*(A) = -\mc L^{-1} \left(\!\!\begin{array}{c}\frac{\theta\xi \gamma\alpha A^2}{\mu_{F_B} +  \alpha A}\\0\\0\\0\\0\end{array}\!\!\right) =:  -\mc L^{-1} \left(\!\!\begin{array}{c}f\\0\\0\\0\\0\end{array}\!\!\right),
$$
and  the coordinates of 
\begin{equation}
b(A) := (F_B^*(A), \mb y^*(A))
\label{bA}
\end{equation}
are increasing functions of $A$. In particular,
$$
F_B^*(A)\leq\lim\limits_{A\to \infty} \frac{\theta \xi \gamma A}{\mu_{F_B} +  \alpha A} = \frac{\theta \xi \gamma}{\alpha} = \frac{\theta \mu_{M_B}}{S}.
$$
To derive the equation for $A^*$, we need parameters related to $\mc L^{-1}$. First,
\begin{align*}
\sf L&:=\text{det} \mc L = T_{R_H}T_{R_V}T_{Q_H}T_{Q_V}T_B(\mc R-1), \end{align*}
 where
 $$
 \mathcal R := \frac{b}{b+\mu_{B}} \left(\frac{(1- \omega) q a_V}{a_V+\mu_{R_V}} \frac{\tau_V V}{\tau_V V+\mu_{Q_V}}  +  \frac{\omega p a_H}{a_H+\mu_{R_H}} \frac{\tau_H H}{\tau_H H+\mu_{Q_H}} \right) =
 \frac{b}{T_{B}} \left(\frac{(1-\omega) q a_V}{T_{R_V}} \frac{\sigma_V}{T_{Q_V}}  +  \frac{\omega p a_H}{T_{R_H}} \frac{\sigma_H}{T_{Q_H}} \right)
 $$
 gives the average number of reproducing females produced from one breeding female. We observe that in this model, we have  $\mathcal R<1$.
Then,
\begin{equation}
\label{RHV}
\begin{split}
R_H&= -f \frac{bT_{R_V}T_{Q_V}\omega p a_H\sigma_H}{a_H\sf L} =
f \frac{b\omega p a_H\sigma_H}{a_HT_{R_H}T_{Q_H}T_B(1-\mc R)}
\\
R_V&= -f \frac{bT_{R_H}T_{Q_H}(1-\omega) q a_V\sigma_V}{a_V\sf L}
=
f \frac{b(1-\omega) q a_V\sigma_V}{a_VT_{R_V}T_{Q_V}T_B(1-\mc R)}.
\end{split}
\end{equation}
Hence,
\begin{equation}
a_HR_H+a_VR_V = {f}\frac{\mc R}{1-\mc R},
\label{ahrh}
\end{equation}
and further
\begin{equation}
B =     f\frac{1}{(b+\mu_B) (1-\mc R)}, \quad Q_H= f\frac{b\omega}{T_{Q_H}T_{B}(1-\mc R)}, \quad Q_V= f\frac{b(1-\omega)}{T_{Q_V}T_{B}(1-\mc R)}.
\label{Betc}
\end{equation}
 We shall need
 $$
 \rho\mathcal D := \frac{\lambda_0\theta\xi \gamma}{\mu_{A_2}} \mc D= \frac{\lambda_0\theta\xi \gamma}{\mu_{A_2}} \frac{\mathcal R}{1-\mathcal R},
 $$
 and $$
 \Gamma := \frac{\gamma + \mu_{A1}}{\mu_{A2}}, \quad \Delta := \frac{\mu_{F_B}}{\alpha}.
 $$
   Then, substituting \eqref{ahrh} into the first equation of system \eqref{Yeq}, after some algebra, we obtain the equation determining the equilibrium values of $A$ as
 \begin{equation}
A\left( \left(l{\rho\mathcal D} +1\right) A^2 + \left(\Gamma+\Delta-\rho\mathcal D\right)A +\Gamma\Delta \right) =0.
 \label{queq}
 \end{equation}
     	
To analyze the nonzero equilibrium values, we see that at any non-zero equilibrium of $A$ we have
     $$
     \rho\mathcal D = \frac{A^2+(\Gamma+\Delta)A +\Gamma\Delta}{A(1-lA)}=:F(A), \quad 0<A<L_P.
     $$
          \begin{figure}[H]
\centering
\includegraphics[scale=0.6]{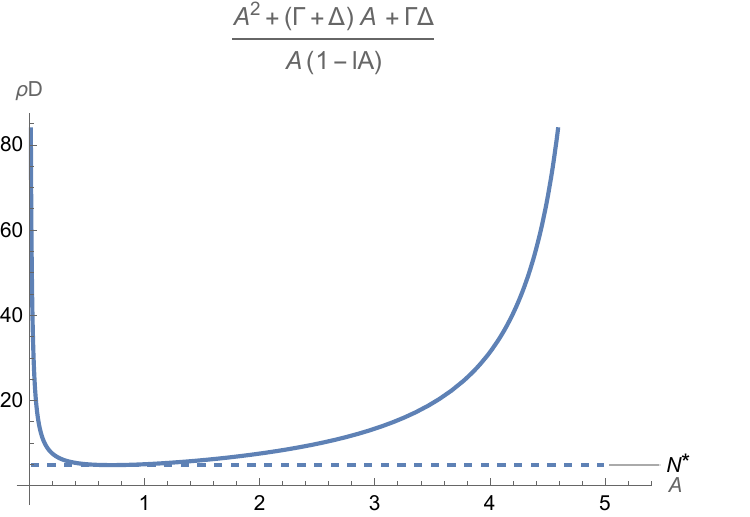}
\caption{Saddle-node bifurcation in the model.} \label{fig:bifurcation}
\end{figure}

     We see that $\lim\limits_{A\to 0,l^{-1}}F(A) = \infty$ and $F(A)$ is monotonically decreasing on $(0,A_{12})$ and monotonically increasing on $(A_{12},l^{-1})$, where the minimum of $F$ is given by 
     $$
     A_{12} = \frac{l\Gamma\Delta + \sqrt{l^2\Gamma^2\Delta^2 + \Gamma\Delta + \Gamma\Delta(1+ (\Gamma+\Delta)l)}}{1+(\Gamma+\Delta) l}.
     $$
     We define
     $$
      N^* = F(A_{12}),
      $$
      and let $A_1$ and $A_2$ be, respectively, the smaller and the larger root of $F(A)=\rho\mc D$ if $\rho\mc D >N^*.$
      
      To find $N^*$, it is easier to use the fact that for $\rho\mc D =N^*$, the quadratic equation in \eqref{queq} has one double root, that is, its discriminant must be zero. This gives
$$
 N^* = \Gamma + \Delta + 2l\Gamma \Delta + 2\sqrt{\Gamma\Delta + l(\Gamma+\Delta)\Gamma\Delta + l^2\Gamma^2\Delta^2}
$$
In the special case $l=0$,
\begin{equation}
 N^* = \left(\sqrt{\Gamma} + \sqrt{\Delta}\right)^2, \quad
A_{12} = \sqrt{\Gamma\Delta}.
\label{N*}
\end{equation}
We define 
$$
\mb x(A) = (A,b(A)).
$$
We can now summarize our considerations in the following theorem. 
	\begin{thm}[Existence of Steady State Solutions]
	
	    \begin{enumerate}
	        \item If $\rho\mathcal D< {N}^*$, then there is only a trivial steady state.
	        \item If $\rho\mathcal{D}={N}^*$, then the trivial steady state coexists with one strictly positive steady state $\boldsymbol x(A_{12})\gg \boldsymbol 0$.
         \item If $\rho\mathcal{D}
         >{N}^*$, then there are the trivial steady state and two strictly positive steady states $\boldsymbol x(A_2)\gg \boldsymbol x(A_{1})\gg \boldsymbol 0$.
         	    \end{enumerate}
Furthermore, for $\rho\mc D>N^*$, $A_1(\rho\mc D)$ monotonically decreases from $A_{12}$ to $0$ and $A_2(\rho\mc D)$ monotonically increases from $A_{12}$ to $L_P$ as $\rho\mc D$ increases from $N^*$ to $\infty$. Thus, the smaller equilibrium 
$\mb x(A_1)$ monotonically decreases to $\mb 0$ and the larger equilibrium $\mb x(A_2)$ monotonically increases to $(L_p,b(L_P))$ as $\rho\mc D\to \infty.$
	    \label{thm:existence}
	\end{thm}
	Next, we shall focus on the stability of the steady states. To continue, we define the point
\begin{equation}\label{eta}
    \boldsymbol{\eta}= \left(L_P, \frac{\theta \mu_{B}}{S},B^0,Q_H^0,Q_V^0,R_H^0,R_V^0 \right) = \left(L_P, \frac{\theta \mu_{B}}{S},\mb y^0\right),\;\text{with}\; 
\mb y^0 = - \mc L^{-1} \left (\begin{array}{c}\theta\xi\gamma L_P\\0\\0\\0\\0\end{array}\right).
\end{equation}
We see that $\boldsymbol{\eta}\in \Omega$, hence  $\mb[\boldsymbol{0},\boldsymbol{\eta}\mb] \subset \Omega$. 

\begin{thm}\label{thm:globalzero}
Suppose \eqref{Laass} holds and  $\rho\mathcal{D} < {N}^*$, where $N^*$ is defined by \eqref{N*}. Then, the trivial steady state is globally asymptotically stable on  $\Omega$.
\end{thm}
\noindent \textbf{Proof}. Let us write   \eqref{eq:fullsystemmodel1} as
\begin{equation}
     \boldsymbol{x}'=\boldsymbol{f}  (\boldsymbol{x}).
\end{equation}
We have $\boldsymbol{f} \left (\boldsymbol{\eta} \right )=\left(-\left(\gamma+\mu_{A1}+\mu_{A2}L_P \right)L_P,-\frac{\mu_{F_B}\theta \mu_{B}}{S},0,0,0,0,0\right)$. By \eqref{RHV} and \eqref{Betc} with $f = \theta\xi\gamma L_P,$ we have 
$\boldsymbol{\eta} \gg\boldsymbol 0. $ Further,    $\boldsymbol{f} \left (\boldsymbol{\eta} \right ) < \boldsymbol{0}$ and $  \boldsymbol{f} \left (\boldsymbol{0} \right ) = \boldsymbol{0}.$ Since, by assumption and Theorem \ref{thm:existence},  the trivial steady state is the only steady state in $[\boldsymbol{0},\boldsymbol{\eta}],$ we conclude that the trivial steady state is globally and asymptotically stable in $[\boldsymbol{0},\boldsymbol{\eta}],$ see  \cite{anguelov2012mathematical} or \cite[Theorem 8.30]{ban2024}. 

 Next, for any $m\in \mathbb{N}$, we define 
 \begin{equation}
 \boldsymbol{\eta}_m= \left(L_P, \frac{\theta \mu_{B}}{S},m\mb y^0 \right).
 \label{etam}
 \end{equation}
 Then, since 
 $$\theta \xi \gamma L_P + a_{H}R_{H}+a_{V}R_{V}-bB-\mu_{B}B = 0, $$
we get 
$$a_{H}R_{H}+a_{V}R_{V}-bB-\mu_{B}B<0.$$
 Hence, 
 $$
 \theta \xi \gamma L_P +m a_{H}R_{H}+ma_{V}R_{V}-mbB-m\mu_{B}B< 0,
 $$
 and therefore 
 $$
 \mc L \mb \eta_m +  \left (\begin{array}{c}\theta\xi\gamma L_P\\0\\0\\0\\0\end{array}\right)\leq \mb 0.
$$
Thus, $\boldsymbol{f} \left ( \boldsymbol{\eta}_m \right ) \leq \boldsymbol{0}$  for any $m\in \mathbb{N}$ and we therefore  $\boldsymbol{0}$ is globally asymptotically stable on $\Omega.$  \hfil $\square$

We precede the analysis of the $\rho\mc D\geq N^*$ case with the following lemma.

\begin{lem}\label{lem:points}
    Suppose \eqref{Laass} holds and  $\rho\mathcal{D} > {N}^*$. There exist points $\boldsymbol{y},\boldsymbol{z}$ in $\Omega$ such that $\boldsymbol{0}\ll \boldsymbol{y}\ll \boldsymbol{x}^*(A_1)\ll \boldsymbol{z}\ll \boldsymbol{x}^*(A_2)$, $\boldsymbol{f}\left(\boldsymbol{y} \right)< \boldsymbol{0} $ and $\boldsymbol{f}\left(\boldsymbol{z} \right)> \boldsymbol{0} $.
\end{lem}
\noindent \textbf{Proof}. Steady states of \eqref{eq:fullsystemmodel1} are obtained by solving \eqref{Yeq}. Dropping the first equation of \eqref{Yeq}, we get\begin{equation}
\begin{split}
	0& =  \theta \xi \gamma A - \alpha AF_B-\mu_{F_B} F_B,\\
0& = \mathcal{L}\boldsymbol{y}+\boldsymbol{G}.
\label{Yeq1}
\end{split}
\end{equation}
For each $A\in [0,L_P]$, we consider the point  $\boldsymbol{b}(A)$  defined in \eqref{bA}. If $A=A_1$ or $A=A_2$, then $\left(A_i,\boldsymbol{b}(A_i) \right), ~~i=1,2,$ is a steady state of \eqref{eq:fullsystemmodel1}. To ensure that this is not the case, we consider the first equation of \eqref{Yeq} and solve for $A$ the inequality
\begin{equation}\label{amm1}
-AP_2(A) >0,
\end{equation}
where $P_2(A)$ the quadratic polynomial in \eqref{queq}. 
Since $A>0$, we see that 
\begin{enumerate}
    \item  $-A P_2(A) >0$ if and only if $A\in (A_1,A_2)$ and
    \item $-A P_2(A) <0$ if and only if $A\in (0,A_1) \cup (A_2, L_P]$.
\end{enumerate}
To complete the proof, we define $\boldsymbol{y}=\left(y,\boldsymbol{b}(y) \right)$ and $\boldsymbol{z}=\left(z,\boldsymbol{b}(z) \right)$, where $y$ is any point in $ (0,A_1)$ and $z$ is any point in $ (A_1,A_2)$ and apply the fact that $b(A)$ is a strictly increasing function of $A$. \hfil $\square$

\begin{thm}\label{thm:statbletwo}
    Suppose \eqref{Laass} holds. 
    \begin{enumerate}
   \item  Let $\rho\mathcal{D} > {N}^*$. Then 
    $[\boldsymbol{0},\boldsymbol{x}^*(A_1)]$ and $\{\boldsymbol x\in \Omega:\boldsymbol x \geq  \boldsymbol{x}^*(A_1)\} $ are positively invariant and  the steady states $\boldsymbol{0}$ and $\boldsymbol{x}^*(A_2)$ are globally asymptotically stable on, respectively, 
$[\boldsymbol{0},\boldsymbol{x}^*(A_1)]\setminus \{\boldsymbol{x}^*(A_1)\}$ and $\{\boldsymbol x\in \Omega:\boldsymbol x > \boldsymbol{x}^*(A_1)\} $.
\item Let $\rho\mathcal{D} = {N}^*$. Then 
    $[\boldsymbol{0},\boldsymbol{x}^*(A_{12})]$ and $\{\boldsymbol x\in \Omega:\boldsymbol x \geq  \boldsymbol{x}^*(A_{12})\} $ are positively invariant and  the steady states $\boldsymbol{0}$ and $\boldsymbol{x}^*(A_{12})$ are globally asymptotically stable on, respectively, 
$[\boldsymbol{0},\boldsymbol{x}^*(A_{12})]\setminus \{\boldsymbol{x}^*(A_{12})\}$ and $\{\boldsymbol x\in \Omega:\boldsymbol x > \boldsymbol{x}^*(A_{12})\} $.
\end{enumerate}
\end{thm}
\noindent \textbf{Proof}. Consider statement 1. The invariance of $[\boldsymbol{0},\boldsymbol{x}^*(A_1)]$ follows directly from \cite[Theorem 8.30]{ban2024}. Using the same result, we see that $[\boldsymbol{0},\mb \eta_m],$ where $\boldsymbol{\eta}_m$ was defined in \eqref{etam}, is invariant, and we obtain the invariance of $\{\boldsymbol x\in \Omega:\boldsymbol x \geq  \boldsymbol{x}^*(A_1)\}$ by passing with $m$ to $\infty$.

Now, let the points $\boldsymbol{y}$ and $\boldsymbol{z}$  be as constructed in the proof of Lemma \ref{lem:points}. We can make $\mb y$  as close to $A_1$ as we wish, and thus $b(\mb y)\ll \mb x^*(A_1)$ can be arbitrarily close coordinatewise to $\mb x^*(A_1).$ Since $\mb f(\mb y) < \mb 0$, $\mb 0$ is globally asymptotically stable on $[\mb 0,\mb x^*(A_1)).$

Moreover, for sufficiently large $m$, we have $\mb x^*(A_1)\ll \boldsymbol{z}\ll\boldsymbol{x}^*(A_2)\ll \boldsymbol{\eta}_m,$ with $\boldsymbol f(\boldsymbol{\eta}_m)<0$  and $b(\mb z)$ can be chosen arbitrarily close to $\mb x^*$. Thus, $\mb x^*(A_2)$ is globally asymptotically stable on $\{\boldsymbol x\in \Omega:\boldsymbol x \gg \boldsymbol{x}^*(A_1)\}$.

To shorten notation, we denote
$$
\mb x^*(A_1)=\mb x^{*}= (x_1^*,\ldots,x_7^*)
$$
and show that the stability can be extended to the faces 
\begin{equation}\label{faces}
\begin{split}
I_i &= \{\mb x : x_i =x_i^*, 0\leq x_j\leq x^*_j\}\setminus\{\mb x^*\}, \quad j=1,\ldots,7,\\J_i &= \{\mb x \in \Omega : x_i =x_i^*, x_j\geq x^*_j\}\setminus\{\mb x^*\}, \quad j=1,\ldots,7.\end{split}
\end{equation}
We observed that both sets $[\boldsymbol{0},\boldsymbol{x}^*(A_1)]$ and  $\{\boldsymbol x\in \Omega:\boldsymbol x \geq  \boldsymbol{x}^*(A_1)\}$ are invariant. We can directly check that no face \eqref{faces} is invariant. Indeed, consider first $I_1$. Then $f_1(x_1^*,x_6,x_7)<0$ if $0<x_6<x_6^*$ or $ 0<x_7<x_7^*$ and thus the only invariant part of $I_1$ can be 
$x_1=x_1^*, x_6=x_6^*$ and $x_7=x_7^*$. But then 
$f_6(x_4,x^*_6)<0$ or $f_7(x_5,x^*_7)<0$ unless $x_4=x^*_4$ and $x_5=x^*_5$. Hence, the only invariant subset can be $x_1=x_1^*, x_4=x^*_4, x_5=x^*_5, x_6=x_6^*,x_7=x_7^*.$ Then, however, 
$f_4(x_3,x_5^*)<0$ or $f_5(x_3,x_5^*)<0$ unless $x_3=x^*_3.$ Arguing as above, $f_3(x_1^*,x_2,x_3^*, x_6^*,x_7^*)<0$ unless $x_2 = x_2^*.$ But then we are at $x^*.$ The other $I_i's$ and $J_i's, i=1,\ldots,7,$ follow in an analogous way. 

Hence, trajectories originating from the faces must enter the interior of the respective order intervals and thus converge to the equilibria in them.  

To prove 2., we observe that we can repeat the above considerations for   $A_{12}=A_1=A_2.$ 
\hfil $\square$

\begin{cor}
    Under assumptions of Theorem \ref{thm:statbletwo} 1., the steady state  $ \boldsymbol{x}^*(A_{1})$ is unstable. If the assumptions of Theorem \ref{thm:statbletwo} 2. are satisfied, the steady state  $ \boldsymbol{x}^*(A_{12})$  is unstable.
\end{cor}
 \textbf{Proof}. In both cases, the steady states do not attract trajectories of the order interval extending from $\mb 0$ to them.  \hfil $\square$

\begin{figure}[H]
\centering
\includegraphics[scale=0.7]{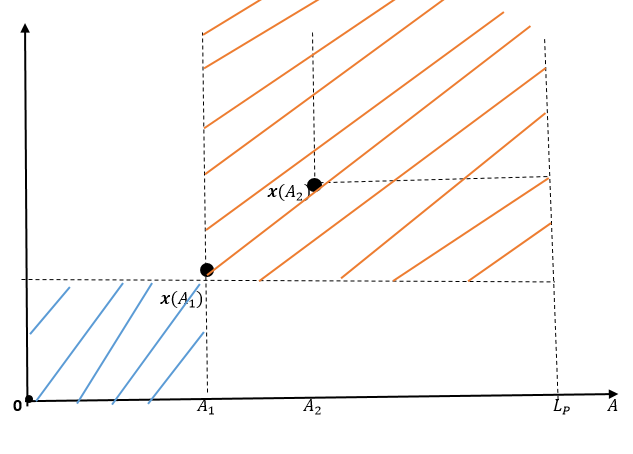}
\caption{2-dimensional representation of the case  ${N} > {N}^*$.
The horizontal axis represents the aquatic lifeforms, while all other coordinates are combined on the vertical axis. The box shaded in orange represents the basin of attraction for the steady state $\boldsymbol{x}^*\left(A_{2} \right) $, while the box shaded in blue represents the basin of attraction of the trivial steady state. } \label{fig:stableregion}
\end{figure}

\section{Asymptotic equivalence of the dynamics}\label{sec4}
To formulate the result on the equivalence of dynamics of \eqref{eq:fullsystem1} and  \eqref{eq:fullsystemmodel1}, we first introduce the necessary notation. 
Let
   $$ \mb x_\epsilon(t) := (A_\epsilon(t),M_{B,\epsilon}(t), F_{B,\epsilon}(t),B_\epsilon(t),Q_{H,\epsilon}(t),Q_{V,\epsilon}(t),R_{H,\epsilon}(t),R_{V,\epsilon}(t))$$
be the solution to \eqref{eq:fullsystem1} with the initial condition $\mb x_\e(0)=\mb{\mr x}$, and
$$
\mb x_0(t) := (A_0(t), F_{B,0}(t),B_0(t),Q_{H,0}(t),Q_{V,0}(t),R_{H,0}(t),R_{V,0}(t)),$$
be the solution to \eqref{eq:fullsystemmodel1}  with $\mb x_0(0) = (\mr A, \mr F_{B},\mr B,\mr Q_{H},\mr Q_{V},\mr R_{H},\mr R_{V}),$ and $$
M_{B,0}(t) = \frac{\xi \gamma}{\mu_{M_B}}A_0(t).
$$
We have 
\begin{thm}
Let 
\begin{align*}
(\mr A, \mr F_{B},\mr B,\mr Q_{H},\mr Q_{V},\mr R_{H},\mr R_{V})&\in\left\{\begin{array}{ccc} \Omega &\text{for} & \mathcal{D}< \rho^{-1}{N}^*,\\
(\mb [\mb 0,\mb x(A^*)\mb ]\setminus\{\mb x(A^*)\}) \cup \{\mb x \in \Omega: \mb x> \mb x(A^*)\}  &\text{for} & \mathcal{D}\geq  \rho^{-1}{N}^*,
\end{array}\right.
\end{align*}
where $A^* = A_1$ if $\mathcal{D} >  \rho^{-1}{N}^*$
and $A^* = A_{12}$ if $\mathcal{D} =  \rho^{-1}{N}^*$, and $\mr M_B\geq 0$ be arbitrary.  
Then 
\begin{equation}\label{tikh1}
(A_\epsilon(t),F_{B,\epsilon}(t),B_\epsilon(t),Q_{H,\epsilon}(t),Q_{V,\epsilon}(t),R_{H,\epsilon}(t),R_{V,\epsilon}(t))=\mb x_0(t)+ O(\epsilon),\quad \e\to 0,\end{equation}
uniformly for $t\in [0,\infty),$
and, for any $t_0>0,$
\begin{equation}\label{tikh2}
M_{B,\epsilon}(t)= M_{B,0}(t) + O(\epsilon),\quad \e\to 0,
\end{equation}
uniformly for $t\in [t_0,\infty).$ 
\end{thm} 
Before the proof, we note that \eqref{tikh2} is valid only on intervals separated from zero because we have not included the initial layer to compensate for the fact that arbitrary initial conditions cannot satisfy \eqref{qss1}. As explained in \cite{BanViet, BanM2AS}, this can be easily remedied, but we skip this part to shorten the discussion.   

\noindent
\textbf{Proof.} The proof is a standard application of \cite[Theorem 1 \& Remark 3]{BanViet}, see also \cite[Appendices A \& B]{BanTch}. It is easy to see that all assumptions of the classical Tikhonov theorem, \cite{TVS}, are satisfied, with \eqref{qss1} defining the only quasi-steady state whose all points are uniformly asymptotically stable equilibria of the fast equation  $\frac{\Tilde M_{B}}{d\tau}  =  \xi \gamma A-\mu_{M_B}\Tilde M_B, \tau = t/\e,$ and any $\mr M_B$ belongs to the basin of attraction of \begin{equation}\label{aux}\frac{\Tilde M_{B}}{d\tau}  =  \xi \gamma \mr A-\mu_{M_B}\Tilde M_B\end{equation} for any $\mr A$ (see assumptions a)--d) of \cite[Appendix A]{BanTch}).  This gives the validity of the approximation \eqref{tikh1} and \eqref{tikh2} on bounded time intervals. To extend it to infinite intervals, we use \cite[Remark 3]{BanViet}, which ensures that it is possible for any $X_\e(t)$ emanating from $(\mr A, \mr M_B,\mr F_{B},\mr B,\mr Q_{H},\mr Q_{V},\mr R_{H},\mr R_{V})$ such that 
$(\mr A, \mr F_B,\mr B,\mr Q_{H},\mr Q_{V},\mr R_{H},\mr R_{V})$ is in the domain of attraction of a (hyperbolic) equilibrium of \eqref{eq:fullsystemmodel1}, and $\mr F_B$ is in the basin of attraction of equilibria of \eqref{aux}, that is, it is arbitrary.  Then, e.g., \cite[Theorem 1]{BanViet} ensures that $X_\e(t)$ converges to $(A_0(t), M_{B,0}(t), F_{B,0}(t),B_0(t),Q_{H,0}(t),Q_{V,0}(t),R_{H,0}(t),R_{V,0}(t))$ uniformly in $t$ on any interval $[0,\infty)$, except for $M_B(t),$ which converges to $M_{B,0}(t)$ uniformly on $[t_0,0)$ for any $t_0>0.$ 

To get the order of convergence in \eqref{tikh1} and \eqref{tikh2}, we use the higher order result, \cite[Theorem 3]{BanM2AS}, and obtain the desired result by retaining only zero-order terms, as explained in \cite[Theorem B.2]{BanTch}. \hfil $\square$

\section{Numerical Simulations}\label{sec5}
We now numerically solve systems  \eqref{eq:fullsystemmodel1} and \eqref{eq:fullsystem1}, and compare the results for different values of $\epsilon$.
The parameter values used are: $\lambda_0 = 5,
a_H = 0.8,
a_V = 0.8,
L_P = 1000,
\gamma = 0.1,
\mu_{A1} = 0.001,
\mu_{A2} = 0.0002,
\theta = 0.5,
\xi = 1,
\mu_{M_B} = \frac{1}{20},
\mu_{F_B} = \frac{1}{20},
b = 1,
\omega = 0.5,
\tau_H = 0.05,
\tau_V = 0.05,
\mu_{Q_H} = \frac{1}{20},
\mu_{Q_V} = \frac{1}{20},
\mu_{R_H} = \frac{1}{20},
\mu_{R_V} = \frac{1}{20},
\mu_B = \frac{1}{20},
p = 0.8,
q = 0.9,
S = 0.01$. With these parameter values, we have three steady-state solutions corresponding to $A_0=0, A_1=4.99212$, and $A_2=142.943$. As we see, our approximations are good even when $\epsilon = 0.3$. $\epsilon = 0.3$ means that the death rate of the male mosquitoes is about three times higher than that of the newly hatched female mosquitoes. This is quite reasonable since male mosquitoes typically live for 6 to 7 days while the females live for 14 to 28 days, \cite{styer2007mosquitoes}.
 
\begin{figure}[H]
\centering
\includegraphics[scale=0.34]{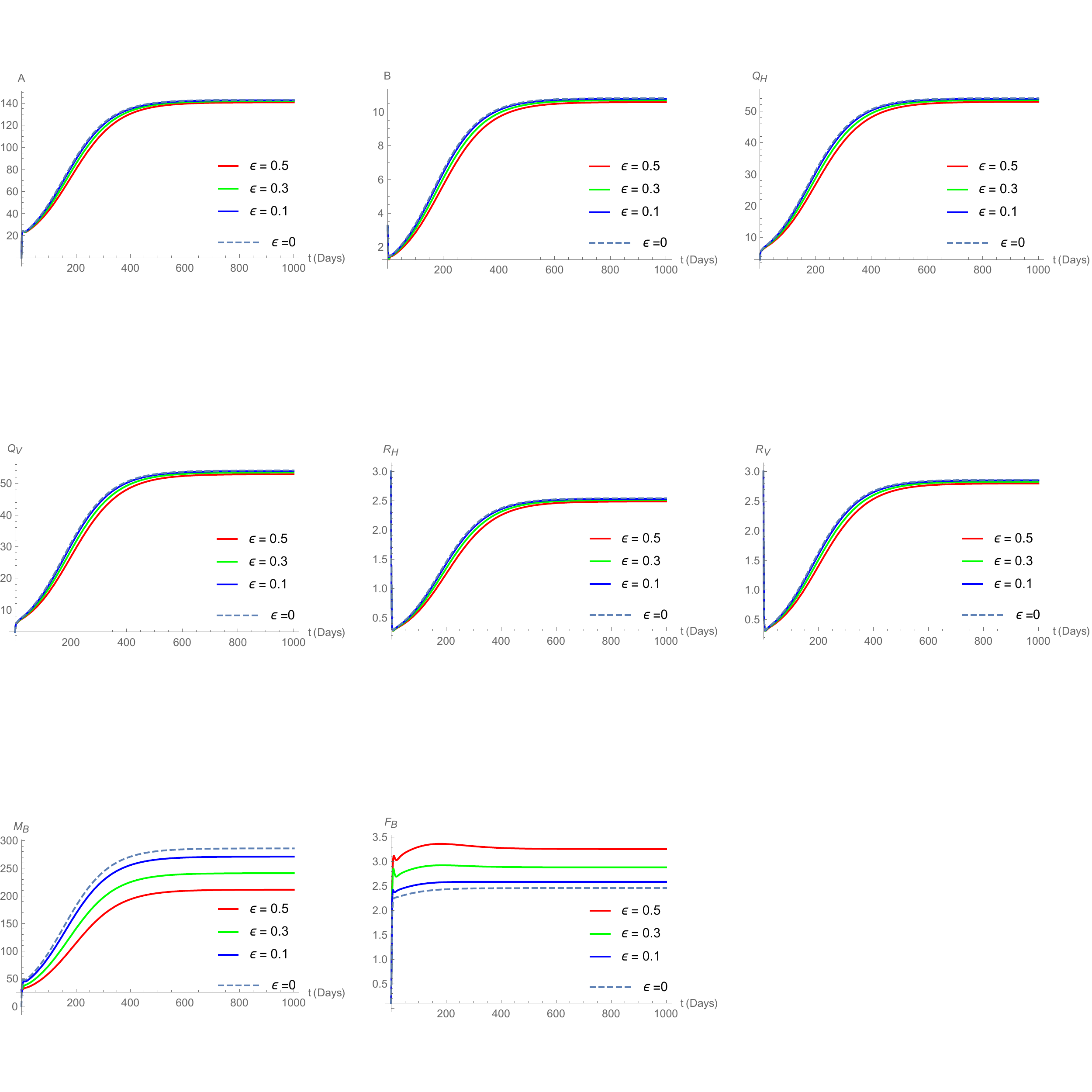}
\caption{  The quality of approximation for initial conditions in the basin of attraction of the stable equilibrium $\mb x(142.943)$ for different values of $\e$; the case $\epsilon=0$ corresponds to the solution of \eqref{eq:fullsystemmodel1}.  }\label{fig:stablenonzero}
\end{figure}

\begin{figure}[H]
\centering
\includegraphics[scale=0.34]{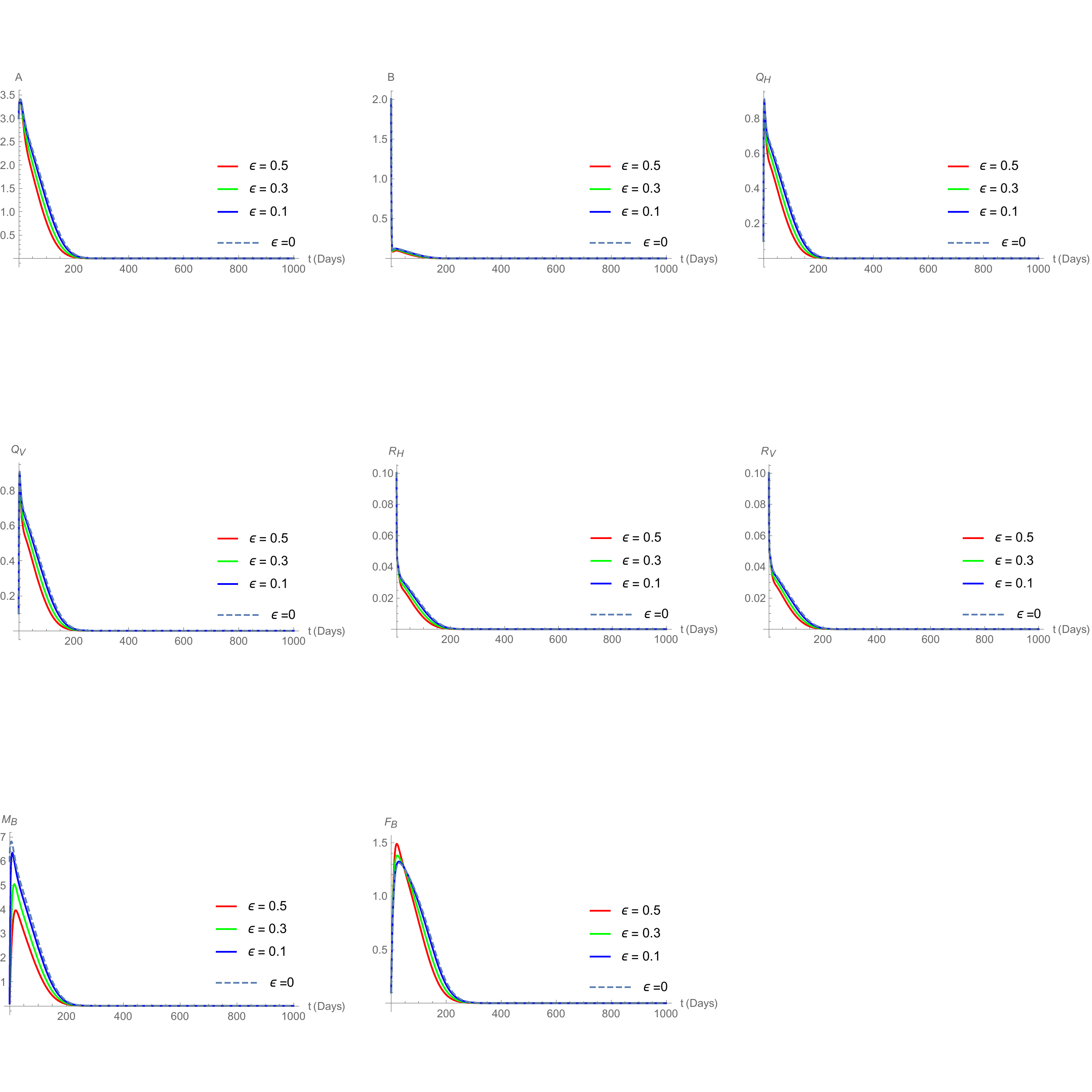}
\caption{The parameter values used here are the same as those in Figure \ref{fig:stablenonzero}. Without altering these parameter values, we modify the initial conditions so that they fall within the basin of attraction of the trivial steady state. The trajectory for $\epsilon=0$ corresponds to the solution of \eqref{eq:fullsystemmodel1}.  }\label{fig:stablezero}
\end{figure}

\section{Conclusions}\label{sec6} 
In this paper, we considered an eight-stage model of a mosquito life cycle introduced in \cite{NMBB}. We used the model's multiscale character to reduce its complexity. Using recent results in asymptotic analysis, we show that the reduced model retains salient features of the original dynamics as long as the initial condition is in the basin of attraction of a (hyperbolic) equilibria of the reduced model. Since the latter is monotone, analysing its dynamics is relatively straightforward. In particular, we identify the bifurcation parameter, which nicely splits into two parts, one containing only post-mating parameters and the other involving only the parameters of the aquatic and the pre-mating stages. If the bifurcation parameter is below the explicitly evaluated threshold, only the trivial equilibrium, which is globally asymptotically stable, exists. When the bifurcation parameter crosses the threshold, we observe a saddle-node bifurcation, which has the additional feature that the two new equilibria are ordered (with respect to the partial order in $\overline{\mbb R}^7_+),$ with the smaller being unstable, and the trivial and the larger asymptotically stable, in the order intervals stretching from $\mb 0$ to the smaller equilibrium and from the smaller equilibrium to the boundary of the feasible domain $\Omega$, respectively. Thus, the dynamics of the model resembles the Allee dynamics. Thus, we can say that the model exhibits a multidimensional Allee effect. In particular, when the bifurcation parameter (for instance, the number of laid eggs per capita  $\la_0$) tends to infinity, the smaller equilibrium tends to zero, destabilizing thus the trivial equilibrium by shrinking its basin of attraction on $\mb 0$. In comparison, the larger equilibrium tends to the boundary of $\Omega$, and thus, the larger equilibrium asymptotically becomes globally stable there. Using asymptotic analysis, we show that these properties are also the features of the dynamics of the original problem.


\bibliographystyle{plain}
\bibliography{tbime}

\begin{thebibliography}{10}

\bibitem{anguelov2012mathematical}
R.~Anguelov, Y.~Dumont, and J.~Lubuma.
\newblock Mathematical modeling of sterile insect technology for control of
  anopheles mosquito.
\newblock {\em Computers \& Mathematics with Applications}, 64(3):374--389,
  2012.

\bibitem{BanM2AS}
J.~Banasiak.
\newblock Some remarks on the renormalization group and {C}hapman-{E}nskog type
  methods in singularly perturbed problems.
\newblock {\em Math. Methods Appl. Sci.}, 43(18):10361--10380, 2020.

\bibitem{BanViet}
J.~Banasiak.
\newblock A note on the {T}ikhonov theorem on an infinite interval.
\newblock {\em Vietnam J. Math.}, 49(1):69--86, 2021.

\bibitem{ban2024}
J.~Banasiak.
\newblock {\em Introduction to Mathematical Methods in Population Theory}.
\newblock Springer Verlag, 2024.

\bibitem{BBN}
J.~Banasiak, B.~M. Ghakanyuy, and G.A. Ngwa.
\newblock Emergence and analysis of multidimensional {A}llee effect in mosquito
  life cycle.
\newblock in preparation.

\bibitem{BL}
J.~Banasiak and M.~Lachowicz.
\newblock {\em Methods of small parameter in mathematical biology}.
\newblock Springer, Heidelberg/New York, 2014.

\bibitem{BanTch}
J.~Banasiak and S.Y. Tchoumi.
\newblock Multiscale malaria models and their uniform in-time asymptotic
  analysis.
\newblock {\em Mathematics and Computers in Simulation}, 221:1--18, 2024.

\bibitem{CapMBS}
E.~Beretta, V.~Capasso, and D.~G. Garao.
\newblock A mathematical model for malaria transmission with asymptomatic
  carriers and two age groups in the human population.
\newblock {\em Math. Biosci.}, 300:87--101, 2018.

\bibitem{clements1992biology}
A.~N. Clements.
\newblock {\em The biology of mosquitoes, Volume 1: Development, nutrition and
  reproduction}.
\newblock Cabi GB, 2023.

\bibitem{dumont2016human}
Y.~Dumont and J.~Thuilliez.
\newblock Human behaviors: A threat to mosquito control?
\newblock {\em Mathematical biosciences}, 281:9--23, 2016.

\bibitem{NMBB}
B.~M. Ghakanyuy, M.~I. Tebogh-Ewungkem, J.~Banasiak, and G.~A. Ngwa.
\newblock Mating versus alternative blood sources as determinants to mosquito
  abundance and population resilience.
\newblock in preparation.

\bibitem{ghakanyuy2022investigating}
B.~M. Ghakanyuy, M.~I. Teboh-Ewungkem, K.~A. Schneider, and G.~A. Ngwa.
\newblock Investigating the impact of multiple feeding attempts on mosquito
  dynamics via mathematical models.
\newblock {\em Mathematical Biosciences}, 350:108832, 2022.

\bibitem{goindin2015parity}
D.~Goindin, Ch. Delannay, C.~Ramdini, J.~Gustave, and F.~Fouque.
\newblock Parity and longevity of aedes aegypti according to temperatures in
  controlled conditions and consequences on dengue transmission risks.
\newblock {\em PloS one}, 10(8):e0135489, 2015.

\bibitem{Hek}
G.~Hek.
\newblock Geometric singular perturbation theory in biological practice.
\newblock {\em Mathematical Biology}, 60:347--386, 2010.

\bibitem{Hopp}
F.~Ch. Hoppensteadt.
\newblock Singular perturbations on the infinite interval.
\newblock {\em Trans. Amer. Math. Soc.}, 123:521--535, 1966.

\bibitem{Kue}
Ch. Kuehn.
\newblock {\em Multiple time scale dynamics}, volume 191 of {\em Applied
  Mathematical Sciences}.
\newblock Springer, Cham, 2015.

\bibitem{li}
J.~Li.
\newblock Discrete-time models with mosquitoes carrying genetically-modified
  bacteria.
\newblock {\em Mathematical Biosciences}, 240(1):35--44, 2012.

\bibitem{ngwa2019three}
G.~A. Ngwa, M.~I. Teboh-Ewungkem, Y.~Dumont, R.~Ouifki, and J.~Banasiak.
\newblock On a three-stage structured model for the dynamics of malaria
  transmission with human treatment, adult vector demographics and one aquatic
  stage.
\newblock {\em Journal of {T}heoretical {B}iology}, 481:202--222, 2019.

\bibitem{RashKooi}
P.~Rashkov and B.~W. Kooi.
\newblock Complexity of host-vector dynamics in a two-strain dengue model.
\newblock {\em J. Biol. Dyn.}, 15(1):35--72, 2021.

\bibitem{RashVent}
P.~Rashkov, E.~Venturino, M.~Aguiar, N.~Stollenwerk, and B.~W. Kooi.
\newblock On the role of vector modeling in a minimalistic epidemic model.
\newblock {\em Math. Biosci. Eng.}, 16(5):4314--4338, 2019.

\bibitem{RASS}
F.~Rocha, M.~Aguiar, M.~Souza, and N.~Stollenwerk.
\newblock Time-scale separation and centre manifold analysis describing
  vector-borne disease dynamics.
\newblock {\em International Journal of Computer Mathematics},
  90(10):2105--2125, 2013.

\bibitem{smith1995theory}
H.~L Smith and P.~Waltman.
\newblock {\em The theory of the chemostat: dynamics of microbial competition},
  volume~13.
\newblock Cambridge {U}niversity {P}ress, 1995.

\bibitem{styer2007mosquitoes}
L.~M. Styer, J.~R. Carey, J.-L. Wang, and T.~W. Scott.
\newblock Mosquitoes do senesce: departure from the paradigm of constant
  mortality.
\newblock {\em The American journal of tropical medicine and hygiene},
  76(1):111, 2007.

\bibitem{TVS}
A.N. Tikhonov, A.B. Vasileva, and A.G. Sveshnikov.
\newblock {\em Differential equations}.
\newblock Springer, Berlin, 1985.

\bibitem{VDCI}
{Vector Disease Control International}.
\newblock Mosquito biology 101: Life cycle.
\newblock \url{https://www.vdci.net/mosquito-biology-101-life-cycle/}, 2024.
\newblock Accessed: \today.

\end{thebibliography}
\end{document}